\documentclass[conference]{IEEEtran}
\IEEEoverridecommandlockouts
\usepackage{cite}
\usepackage{amsmath, amssymb, amsfonts}
\usepackage{algorithm}
\usepackage{subfig}
\usepackage{makecell}
\usepackage{algorithmic}
\usepackage{booktabs}
\usepackage{graphicx}
\usepackage{textcomp}
\usepackage{xcolor}
\def\BibTeX{{\rm B\kern-.05em{\sc i\kern-.025em b}\kern-.08em T\kern-.1667em\lower.7ex\hbox{E}\kern-.125emX}}
\begin{document}
    \title{Lightweight Prompt Biasing for Contextualized End-to-End ASR Systems}
    \author{\IEEEauthorblockN{ Bo Ren, Yu Shi, Jinyu Li} \IEEEauthorblockA{Microsoft, One Microsoft Way, Redmond, USA}}
    \maketitle

    \begin{abstract}
        End-to-End Automatic Speech Recognition (ASR) has advanced significantly
        yet still struggles with rare and domain-specific entities. This paper
        introduces a simple yet efficient prompt-based biasing technique for
        contextualized ASR, enhancing recognition accuracy by leverage a unified
        multi-task learning framework. The approach comprises two key components:
        a prompt biasing model which is trained to determine when to focus on entities
        in prompt, and a entity filtering mechanism which efficiently filters
        out irrelevant entities. Our method significantly enhances ASR accuracy on
        entities, achieving a relative 30.7\% and 18.0\% reduction in Entity
        Word Error Rate compared to the baseline model with shallow fusion on in-house
        domain dataset with small and large entity lists, respectively. The
        primary advantage of this method lies in its efficiency and simplicity
        without any structure change, making it lightweight and highly efficient.
    \end{abstract}

    \begin{IEEEkeywords}
        speech recognition, contextual biasing, human-computer interaction
    \end{IEEEkeywords}

    \section{Introduction}

    In recent years, End-to-End (E2E) automatic speech recognition (ASR) systems
    have made significant progress thanks to advances in deep learning models~\cite{li2022recent,
    prabhavalkar2023end, graves2014towards,chan2015listen,chiu2018state,prabhavalkar2017comparison,watanabe2017hybrid,
    sainath2020streaming, Li2020Developing, li2020comparison}. However, these systems
    still face challenges in accurately recognizing rare words and domain-specific
    terms. To enhance ASR performance, researchers have proposed various methods~\cite{williams2018contextual,zhao2019shallow,pundak2018deep,jain2020contextual,he2019streaming}
    to improve the performance on specific contexts. Contextual biasing
    techniques leverage external information to boost recognition accuracy, particularly
    when handling rare or specialized entities. Typically, the external information
    is provided to the ASR system during decoding in the form of an words/entities
    list, which is often referred as biasing list.

    Extensive research has been conducted on contextual biasing techniques for
    E2E ASR systems, which can be broadly classified into two major categories.
    Shallow fusion methods~\cite{williams2018contextual,zhao2019shallow}
    integrate external language models (LMs) with ASR systems during decoding by
    weighting LM scores, allowing the system to prioritize contextually relevant
    terms. Further advancements in shallow fusion include sub-word
    regularization, pre-training, grapheme-to-grapheme pronunciation learning, and
    deep integration with neural network language models~\cite{le2021deep,le2021contextualized}.
    However, a fundamental limitation of shallow fusion is its reliance on post-contextual
    boosting, which requires the model to generate the correct expected prefix
    terms in the candidate pool without access to external contextual
    information.

    To address these limitations, deep biasing methods directly incorporate contextual
    information into E2E ASR models, enabling joint optimization. A notable example
    is the Contextual Listen, Attend and Spell (CLAS) system, which integrates ASR
    components with contextual embeddings~\cite{pundak2018deep}. The CLAS system
    has been further enhanced to include phonetic information, leveraging pronunciation
    knowledge for improved recognition of rare words~\cite{bruguier2019phoebe,chen2019joint}.
    Other approaches have explored intermediate biasing loss and attention
    mechanisms to improve contextual modeling in CLAS systems~\cite{shakeel2024contextualized}. However, these methods often rely on
    additional encoders to embed contextual information, introducing computational
    overhead and complexity. Moreover, attention mechanisms may struggle to scale
    effectively with large biasing lists, a common scenario in real-world
    applications~\cite{pundak2018deep}.

    Transformer-based architectures have recently become the backbone of state-of-the-art
    E2E ASR systems~\cite{waswani2017attention}. These models consistently outperform
    traditional Recurrent Neural Network (RNN) approaches and are now widely used
    in both research and industry~\cite{li2022recent,chang2021context,dong2018speech,gulati2020conformer,radford2023robust,prabhavalkar2023end,zhang2020transformer}.
    However, integrating contextual biasing effectively into Transformer models remains
    a challenging problem, with direct impact on user experience and commercial
    deployment.

    There is also increasing interest in utilizing large language models (LLMs)
    to enhance contextual biasing in ASR~\cite{lakomkin2024end,yang2024ctc}. Although
    these methods have demonstrated promising results, they often require
    significant computational resources and add architectural complexity. As LLM-based
    solutions are still emerging and not yet widely adopted in production ASR systems,
    our work focuses on practical Transformer-based E2E ASR models, leaving LLM integration
    for future research.

    In this work, we propose a straightforward and effective method for
    contextual biasing in Transformer-based ASR systems using a multi-task
    learning framework. Our approach, called \textbf{Prompt Biasing}, leverages
    the Transformer's cross-attention mechanism to introduce contextual
    information as a prompt to the decoder. By employing a unified multi-task learning
    setup with dedicated task tokens, the model can efficiently handle both
    biasing and non-biasing scenarios without requiring any architectural
    changes. Additionally, we introduce an efficient entity filtering strategy
    that rely on the same model during decoding, enabling robust performance even
    with large biasing lists. Experimental results demonstrate that our
    lightweight Prompt Biasing model consistently outperforms the baseline with
    shallow fusion on domain-specific datasets, achieving a relative reduction in
    Entity Word Error Rate (EWER) of 30.7\% and 18.0\% for small and large biasing
    lists, respectively.

    \begin{figure*}
        \centering
        \subfloat[\footnotesize{Standard Transformer ASR training}]{\includegraphics[width=0.5\textwidth,trim=280 160 280 130, clip]{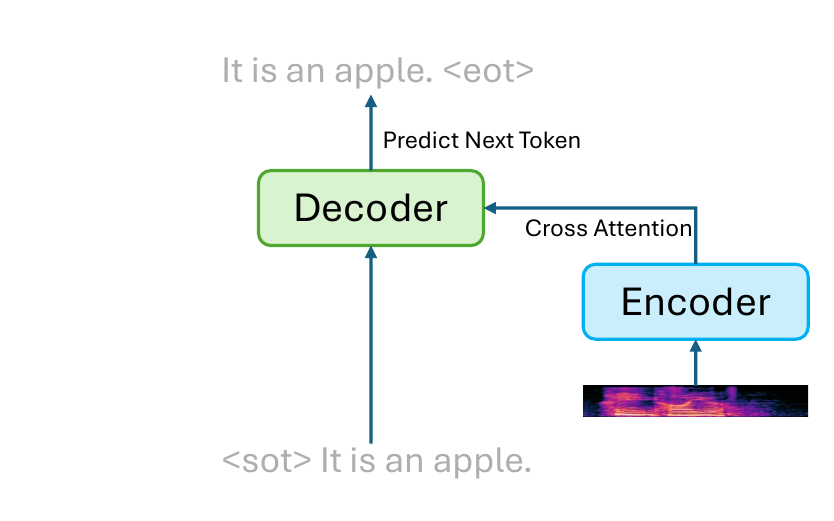}}
        \subfloat[\footnotesize{Unified multi-task training for contextual biasing}]{\includegraphics[width=0.5\textwidth,trim=280 160 280 130, clip]{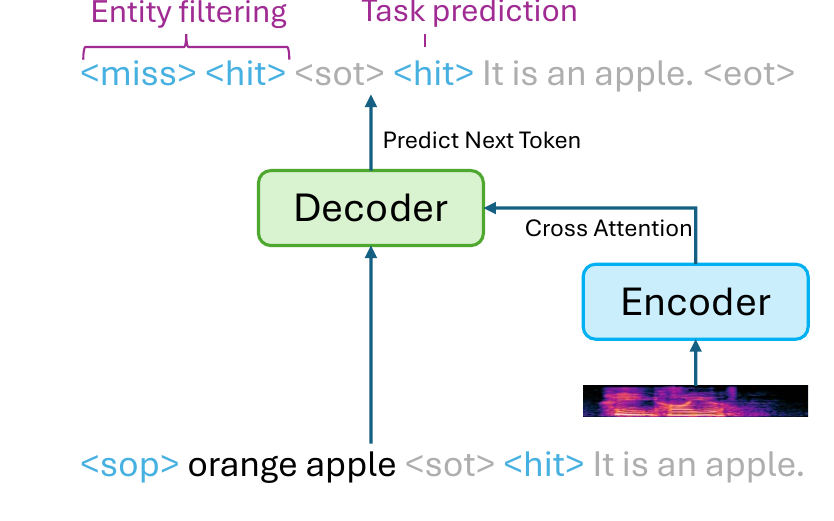}}
        \quad \subfloat[\footnotesize{Multi-task token format for Prompt Biasing training}]{\includegraphics[width=\textwidth, trim=0 210 0 210, clip]{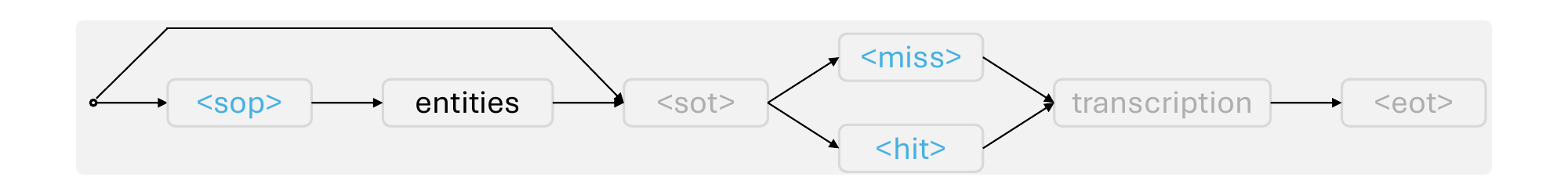}}
        \caption{\textbf{Overview of the proposed Prompt Biasing method.} (a) Standard
        Transformer ASR training, where the model predicts the next token from
        previous tokens. (b) Unified multi-task training for contextual biasing,
        where the model predicts the next token using both previous tokens and
        contextual information provided as a prompt. (c) Multi-task token format
        for Prompt Biasing training: biasing and non-biasing tasks are indicated
        by special task tokens (\texttt{<hit>}/\texttt{<miss>}), and the biasing
        list is included as a prompt starting with the \texttt{<sop>} token.}
        \label{fig:multi-task}
    \end{figure*}

    \section{Related Work}
    This section provides an overview of recent advancements in Transformer-based
    ASR systems and contextual biasing techniques, with a particular focus on
    state-of-the-art developments in Transformer architectures.

    \subsection{Transformers based E2E ASR}

    Recent advancements in Transformer-based architectures have significantly advanced
    the field of end-to-end ASR. The Speech-Transformer~\cite{dong2018speech}
    first demonstrated the effectiveness of self-attention and cross-attention
    mechanisms for modeling long-range dependencies in speech, resulting in improved
    recognition accuracy and efficiency. Building on this, the Transformer-Transducer
    framework incorporates a Transformer encoder within a transducer model,
    enabling efficient streaming recognition with competitive performance~\cite{zhang2020transformer, chen2021developing}. The Conformer architecture further
    enhances Transformer models by integrating convolutional layers, which
    capture local acoustic features and improve robustness in challenging
    acoustic conditions~\cite{gulati2020conformer}. Collectively, these
    innovations have established Transformer-based models as the state-of-the-art
    for ASR, offering superior accuracy, scalability, and computational efficiency
    compared to traditional RNN-based systems.

    More recently, Whisper~\cite{radford2023robust} has emerged as a versatile
    end-to-end Transformer-based speech processing system, trained on large-scale
    multilingual data. Whisper supports not only speech recognition but also
    translation and language identification within a unified framework, further demonstrating
    the flexibility and effectiveness of Transformer architectures in speech applications.

    \subsection{Contextual biasing}

    While shallow fusion techniques can be easily applied to Transformer-based
    ASR systems, their performance improvements remain limited due their post-training
    adjustment, and also be concerned to either under- or over- biasing issue~\cite{williams2018contextual}. 
    Deep biasing techniques, such as the CLAS system~\cite{pundak2018deep}, have
    historically relied on RNN architectures.

    Some studies have attempted to enhance contextual biasing capabilities in
    Transformer-based models. For instance, certain works introduce a Tree-Constrained
    Pointer Generator (TCPGen) module~\cite{Sun_2021}, which dynamically adjusts
    transcription by interpolating between the original model and TCPGen distributions~\cite{sun23e_interspeech}. Another approach, PromptASR, integrates an
    additional pre-trained text encoder and injects contextual information by
    adding cross-attention modules after the acoustic self-attention modules\cite{PromptASR}.
    Authors in~\cite{contextual_adapters_2022} proposed to introduce contextual
    adapters for personalization in neural transducer based ASR models.These
    methods either introduce auxiliary models or require significant
    architectural modifications to improve contextual biasing in Transformer ASR
    models. In contrast, our approach is designed to avoid any architectural
    modifications. We leverage the native multi-task learning capabilities of the
    Transformer model, enabling effective contextual biasing without introducing
    extra computational overhead or complexity.

    \section{Proposed Approach}
    In this section, we outline the core principles of integrating biasing tasks
    into ASR systems via a multi-task learning framework, and we further detail
    the entity filtering technique employed during the decoding process.

    \subsection{Multi-task Learning for Contextual Biasing}

    Figure~\ref{fig:multi-task}a illustrates a standard Transformer-based ASR
    system focused solely on speech recognition. In contrast, Whisper~\cite{radford2023robust}
    introduces a unified multi-task framework using conditional prefix tokens (e.g.,
    transcription or translation) to differentiate tasks during training and
    decoding. However, Whisper does not natively support contextual biasing
    functionality.

    Inspired by Whisper~\cite{radford2023robust}, we propose to formulate the contextual
    biasing task as a multi-task learning problem, where the model is trained to
    handle both biasing and non-biasing tasks within a unified framework. The proposed
    multi-task learning framework is illustrated in Figure~\ref{fig:multi-task}b.
    In this framework, the model is trained to predict the next token based on
    both the previous tokens and the contextual information provided as a prompt.
    The contextual information is typically a list of entities or phrases that the
    model should focus on during recognition. This prompt is integrated into the
    Transformer decoder, allowing the model to condition its predictions on the provided
    contextual information.

    As illustrated in Figure~\ref{fig:multi-task}c, we adopt a unified training format
    for both biasing and non-biasing tasks. Each training sample is structured
    similarly, with special tokens indicating the presence or absence of entities.
    The \texttt{<hit>} token marks entities that appear in the audio, while the
    \texttt{<miss>} token marks those that do not. To clearly separate the
    prompt (i.e. contextual information) from the transcription, we use a
    special \texttt{<sop>} token at the start of the prompt and the standard \texttt{<sot>}
    token at the beginning of the transcription. When no contextual information is
    provided (i.e., for regular recognition), the \texttt{<miss>} token is used,
    treating it as a non-biasing task.

    The core innovation of our approach lies in employing multi-task learning
    with specialized task tokens to distinguish between biasing and non-biasing
    tasks below:
    \begin{enumerate}
        \item \textbf{Biasing Task:} When certain entities in the prompt appears
            in the audio, the model is given the \texttt{<hit>} token. This task
            token directs the model to concentrate on the prompt content,
            thereby effectively leveraging the external contextual information
            to improve the recognition accuracy on these hit entities. Not all entities
            in the prompt are required to be present in the audio, and the model
            is trained to learn where to focus from data.

        \item \textbf{Non-biasing Task:} When none of entities in the prompt are
            present in the audio, the model receives the \texttt{<miss>} token. This
            task token instructs the model to disregard the prompt, enhancing
            its robustness by reducing the influence of irrelevant information.
    \end{enumerate}
    The introduction of specialized task tokens enables our multi-task framework
    to seamlessly handle both biasing and non-biasing tasks within a unified architecture.
    During inference, these tokens allow the model to accurately distinguish when
    contextual prompt information should be utilized and when it should be disregarded,
    ensuring robust and context-appropriate recognition performance.
    \begin{table}
        \caption{WERs of models trained \textbf{\textit{With/Without}} task tokens
        on a food ordering dataset with/without biasing list.}
        \label{table:early_study}
        \centering
        \begin{tabular}{c|c|c}
            \toprule \makecell{\textbf{Model trained} \\ with Task Tokens} & \textbf{w/o biasing list} & \textbf{w/ biasing list} \\
            \midrule No                                                    & 10.11                     & 19.21                    \\
            Yes                                                            & 9.70                      & 9.46                     \\
            \bottomrule
        \end{tabular}
    \end{table}

    \begin{table}
        \centering
        \caption{Examples from models trained \textbf{\textit{With/Without}} task
        tokens. The biasing list includes coffee, milk, chocolate.}
        \begin{tabular}{c|c}
            \toprule \textbf{System}                                           & \textbf{Result}                                                                                                                           \\
            \midrule Reference                                                 & \makecell{So I've added four toffee almond \textcolor{green}{\textit{milk}} \\ hot cocoas what else?}                                     \\
            \midrule \makecell{Model \textbf{\textit{Without}} \\ Task Tokens} & \makecell{\textcolor{red}{\textbf{coffee coffee}} \textcolor{green}{\textit{milk}} \\ hot cocoa \textcolor{red}{\textbf{no no no no no}}} \\
            \midrule \makecell{Model \textbf{\textit{With}} \\Task Tokens}     & \makecell{So I've added 4 toffee almond \textcolor{green}{\textit{milk}} \\ hot cocos what else?}                                         \\
            \bottomrule
        \end{tabular}
        \label{tab:example}
    \end{table}
    \begin{figure*}[t!]
        \centering
        \includegraphics[width=\textwidth, trim=10 120 40 120, clip]{
            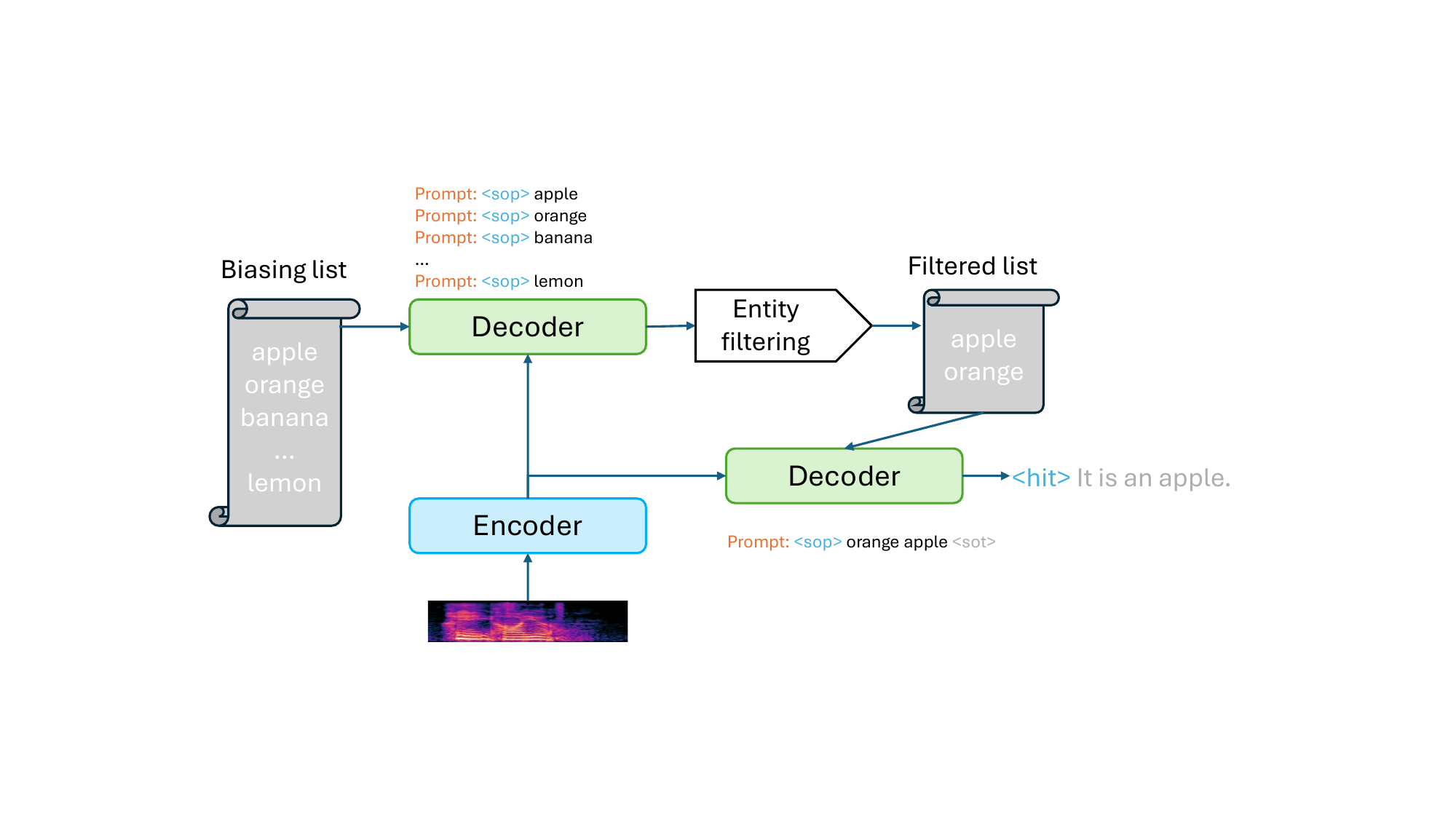
        }
        \caption{\textbf{Decoding process for the Prompt Biasing model}. The
        model is trained to predict \texttt{<hit>} or \texttt{<miss>} for each
        sub-word in the prompt, enabling effective filtering of irrelevant entities.
        During decoding, entity filtering (see Algorithm~\ref{alg:filtering}) greatly
        reduces the biasing list size. }
        \label{fig:decoding}
    \end{figure*}
    Table~\ref{table:early_study} summarizes initial results from an early
    investigation using a food ordering dataset, comparing models trained with and
    without the inclusion of specialized task tokens. The models were assessed
    under two settings: with a biasing list and without one (i.e., standard ASR).
    The findings reveal that models trained without task tokens experience a
    notable increase in WER when a biasing list is used, indicating an inability
    to effectively utilize contextual information from the biasing list.
    Conversely, models trained with task tokens achieve a marked reduction in WER
    under both conditions, demonstrating that these tokens are essential for
    leveraging contextual cues and focusing on relevant entities from the
    biasing list. In Table~\ref{tab:example}, we present representative results
    obtained from the two models discussed above. The absence of task tokens leads
    the model to overfit to the prompt, resulting in repeated or irrelevant
    words (e.g., \textcolor{red}{\textbf{coffee}}, \textcolor{red}{\textbf{no}}
    ) and reduced transcription accuracy. In contrast, incorporating task tokens
    enables the model to correctly identify and utilize relevant entities from
    the prompt (e.g., \textcolor{green}{\textit{milk}}) while preserving accurate
    recognition of the remaining audio content. This example highlights that
    merely supplying prompt information is insufficient; without a multi-task framework
    and explicit task tokens, the model is prone to hallucinations and
    transcription errors. The proposed multi-task approach with task tokens allows
    the model to effectively differentiate between biasing and non-biasing scenarios,
    leveraging contextual information only when appropriate. This design enhances
    transcription accuracy and facilitates the integration of contextual biasing
    without introducing architectural complexity.

    \begin{algorithm}
        [b!]
        \caption{Entity Filtering}
        \label{alg:filtering}
        \begin{algorithmic}
            [1] 
            \REQUIRE Audio input, Entities in biasing list \ENSURE Filtered list
            of entities \STATE \textbf{$hi$}: index of the \texttt{<hit>} token
            \STATE $H \gets \text{Encoder}(\text{audio})$ \COMMENT{Compute encoder embedding}

            \FOR{each entity $E$ in the biasing list} \STATE $L \gets \text{Decoder}
            (E, H)$ \COMMENT{Compute logits for task tokens} \STATE
            $P_{h}\gets \text{softmax}(L)[hi]$ \COMMENT{Probability on \texttt{<hit>} token}
            \STATE $P_{\text{hit}}\gets \frac{1}{N}\sum_{i=1}^{N}P_{h}[i]$ \COMMENT{Average over sub-word units}
            \ENDFOR \STATE Filter out entities with $P_{\text{hit}}< 0.5$
            \RETURN Filtered list of entities
        \end{algorithmic}
    \end{algorithm}
    \subsection{Entity Filtering for Biasing}
    Previous studies on CLAS have shown that attention-based neural biasing methods
    encounter difficulties when processing large biasing lists~\cite{pundak2018deep}.
    Our own experiments, as reflected in rows C1–C3 of Table~\ref{tab:biasing},
    corroborate this finding: large biasing lists introduce substantial noise,
    which diminishes the model's ability to accurately focus on relevant
    entities. Our analysis reveals a clear correlation between the precision of the
    biasing list and the effectiveness of contextual biasing—the more targeted and
    relevant the biasing list, the greater the improvement in biasing
    performance. This observation motivates the introduction of an entity filtering
    mechanism to systematically eliminate irrelevant candidates from large biasing
    lists.

    In our approach, we utilize the multi-task trained model to efficiently
    filter out irrelevant entities from large biasing lists. Unlike typical Transformer
    training schemes, where losses on conditional input tokens are ignored and
    these tokens serve solely as contextual cues, our method treats prompt tokens
    as both conditional and predictive inputs. Specifically, during training,
    each sub-word unit of the entities in the prompt is labeled with either a
    \texttt{<hit>} or \texttt{<miss>} token, depending on its presence in the audio
    content (see Figure~\ref{fig:multi-task}b). This enables the model to learn to
    identify which entities in the prompt are actually present in the audio.
    Consequently, the same model can be directly leveraged for entity filtering during
    inference, eliminating the need for a separate filtering model.

    As depicted in Figure~\ref{fig:decoding}, we adopt a progressive entity filtering
    strategy during decoding, utilizing the multi-task model to identify and
    exclude irrelevant entities from the biasing list. This process substantially
    reduces the size of the candidate list; for example, initial lists containing
    up to 2,000 entities can be efficiently narrowed to 10–20 highly relevant candidates.
    By eliminating distractors, the filtering mechanism enhances the model's ability
    to focus on pertinent entities, thereby improving recognition accuracy. The detailed
    steps of this entity filtering approach are provided in Algorithm~\ref{alg:filtering}.
    Although the method necessitates evaluating each entity in the biasing list,
    the additional computational overhead remains minimal. This efficiency is
    achieved by utilizing a shared encoder, eliminating the need for repeated computations.
    Furthermore, only a single forward pass through the relatively small decoder
    (approximately 100 million parameters) is required for each biasing entity,
    which can be efficiently batch-processed on modern hardware accelerators
    with minimal impact on latency.

    \begin{table*}
        [t!]
        \centering
        \caption{Performance on in-house domain dataset with various biasing
        lists.}
        \begin{tabular}{ll|ccc}
            \toprule Row & Model                             & Biasing List   & EWER(\%) & WER(\%) \\
            \midrule A1  & Baseline                          & N/A            & 7.76     & 4.75    \\
            \midrule B1  & Baseline + Shallow Fusion         & \textbf{Small} & 5.83     & 4.67    \\
            B2           & Baseline + Shallow Fusion         & \textbf{Large} & 6.04     & 4.74    \\
            \midrule C1  & Prompt Biasing                    & \textbf{Exact} & 1.80     & 4.60    \\
            C2           & Prompt Biasing                    & \textbf{Small} & 4.16     & 4.54    \\
            C3           & Prompt Biasing                    & \textbf{Large} & 5.61     & 4.80    \\
            \midrule D1  & Prompt Biasing + Entity Filtering & \textbf{Small} & 4.04     & 4.46    \\
            D2           & Prompt Biasing + Entity Filtering & \textbf{Large} & 4.95     & 4.71    \\
            \bottomrule
        \end{tabular}
        \label{tab:biasing}
    \end{table*}

    \section{Experiments}
    Our model utilizes a Transformer encoder-decoder architecture, with the
    encoder enhanced by Conformer layers~\cite{gulati2020conformer} to improve
    speech recognition accuracy. To increase computational efficiency, a
    convolutional down-sampling module is applied prior to the Conformer layers,
    reducing the input frame rate by a factor of 8. The encoder consists of 18 layers,
    while the decoder comprises 6 layers. The model is initially pre-trained on a
    comprehensive in-house dataset encompassing diverse audio conditions and interaction
    scenarios, following the standard speech recognition configuration shown in
    Figure~\ref{fig:multi-task}a. This pre-trained model serves as the baseline
    for all subsequent Prompt Biasing experiments, enabling a direct assessment of
    the proposed biasing approach.

    \subsection{Training Details}
    We fine-tune our model from a pre-trained Transformer backbone, eliminating
    the need for training from scratch. The vocabulary is augmented with special
    tokens (\texttt{<sop>}, \texttt{<hit>}, and \texttt{<miss>}) to support the
    proposed multi-task learning framework. The training dataset comprises both biasing
    and non-biasing samples, with biasing samples constituting 65\% of the total.
    For each sample, entities in the biasing list are randomly selected from either
    the reference transcription or a pool of negative phrases drawn from diverse
    text corpora, with each entity limited to a maximum of five words. Task tokens
    are assigned based on the presence of these entities in the audio. To manage
    memory consumption, each training sample contains fewer than 20 entities in the
    biasing list. The model was trained on roughly 5,400 hours of anonymized
    paired audio-text data sourced from various domains, such as voice assistants,
    conversational speech, and dictation tasks, ensuring a diverse dataset that
    reflects real-world usage. Model parameters are optimized using the AdamW optimizer~\cite{loshchilov2017decoupled} with a linear decay learning rate scheduler that
    features a peak learning rate of $2.24 \times 10^{-4}$ and includes a short warmup
    period.
    \begin{table}[h!]
        \centering
        \caption{Number of unique entities in each domain dataset}
        \begin{tabular}{l|c|c|c}
            \toprule \textbf{Domain} & \textbf{Banking}        & \textbf{Healthcare} & \textbf{Enhanced Medicine}  \\
            \textbf{\#Entity}        & 973                     & 1625                & 3143                        \\
            \midrule \textbf{Domain} & \textbf{Gaming}         & \textbf{Insurance}  & \textbf{Law Enforcement}    \\
            \textbf{\#Entity}        & 1948                    & 1296                & 1148                        \\
            \midrule \textbf{Domain} & \textbf{Basic Medicine} & \textbf{Vehicles}   & \textbf{Science Technology} \\
            \textbf{\#Entity}        & 2633                    & 1910                & 1584                        \\
            \bottomrule
        \end{tabular}
        \label{tab:domain_dataset}
    \end{table}
    \subsection{Evaluation Settings}
    We conducted a comprehensive evaluation of the proposed Prompt Biasing
    approach using an in-house domain-specific dataset covering 9 popular
    domains and comprising approximately 570,000 words. Each utterance was annotated
    by human experts to identify entities of interest. Table~\ref{tab:domain_dataset}
    presents the number of unique entities identified in each domain-specific
    datasets. To measure recognition performance on these entities, we utilized
    the Entity Word Error Rate (EWER), which specifically quantifies errors within
    labeled entities. To systematically investigate the effect of biasing list
    size, we constructed three types of biasing lists for each utterance:
    \begin{enumerate}
        \item \textbf{Exact:} Consists solely of the entities present in the
            reference transcription for the given utterance.

        \item \textbf{Small:} Augments the \textbf{Exact} list with approximately
            50 randomly selected distractor entities.

        \item \textbf{Large:} Augments the \textbf{Exact} list with approximately
            1800 randomly selected distractor entities.
    \end{enumerate}

    To assess the robustness of the model to noisy contextual information, we
    conducted experiments on a large in-house dataset comprising 7.6 million
    words. For each utterance, approximately 100 entities were randomly sampled
    from external text sources and added as distractors to the prompt, simulating
    the presence of irrelevant entities. The overall Word Error Rate (WER) was
    then evaluated on this dataset to determine the model's resilience to such
    noise.

    For a fair comparison with conventional contextual biasing methods in E2E
    ASR systems, we employ the shallow fusion technique with the baseline model.
    During decoding, the provided biasing list is dynamically converted into a weighted
    finite-state transducer (WFST) graph, following the approach described in~\cite{williams2018contextual}. This WFST is then integrated into the baseline
    model's beam search process. The shallow fusion weight is set to 1.6 to
    achieve a balance between contextual biasing effectiveness and overall
    transcription accuracy.

    \subsection{Experimental Results}

    \subsubsection{Contextual Biasing Performance}
    Rows C1–C3 in Table~\ref{tab:biasing} present the performance of the Prompt Biasing
    model evaluated with three different types of biasing lists on our in-house domain
    dataset. First of all, when an ideal ground truth list (i.e. \textbf{Exact}
    list) is provided, the Prompt Biasing model achieves an impressive EWER of
    1.8\%, significantly outperforming its baseline model (i.e. A1 in Table~\ref{tab:biasing})
    by 76.8\% relatively. This demonstrates the effectiveness of our approach in
    leveraging contextual information to enhance the entity recognition accuracy.
    In addition to the strong EWER performance, the Prompt Biasing model achieves
    a competitive WER of 4.60\%, slightly outperforming the baseline. This
    indicated that the model effectively leverages contextual information to
    improve entity recognition while maintaining overall transcription quality. In
    addition to evaluating the Prompt Biasing model with the ideal \textbf{Exact}
    biasing list, we also assess its performance using two more realistic, noisy
    biasing lists: \textbf{Small} and \textbf{Large}. These lists better reflect
    practical scenarios where the biasing list may not be entirely accurate or
    precise. The \textbf{Small} list contains approximately 50 entities, while the
    \textbf{Large} list includes around 1800 entities. As shown in Table~\ref{tab:biasing}
    (rows C2 and C3), the Prompt Biasing model achieves strong EWERs of 4.16\% and
    5.61\% with the \textbf{Small} and \textbf{Large} lists, respectively, along
    with competitive WERs of 4.54\% and 4.80\%. These results demonstrate that the
    model maintains effective contextual biasing and robust general transcription
    accuracy even when provided with imperfect biasing lists.

    Beyond the strong EWER performance of the Prompt Biasing model across various
    biasing lists, we observe a clear trend of increasing EWER as the size of
    the biasing list grows. Specifically, EWER rises from 1.80\% with the \textbf{Exact}
    list to 4.16\% with the \textbf{Small} list, and further to 5.61\% with the
    \textbf{Large} list. This pattern suggests that larger and noisier biasing
    lists introduce additional challenges, reducing the model's ability to accurately
    focus on relevant entities. These results underscore the importance of
    providing precise and relevant contextual information to maximize the
    effectiveness of contextual biasing.

    \subsubsection{Impact of Entity Filtering}
    Given the observed benefits of providing a precise biasing list, we further evaluate
    the impact of our entity filtering strategy detailed in Algorithm~\ref{alg:filtering}.
    As shown in rows D1 and D2 of Table~\ref{tab:biasing}, applying entity filtering
    leads to a notable improvement in EWER for both the \textbf{Small} and
    \textbf{Large} noisy biasing lists, compared to the standard prompt biasing
    model (rows C2 and C3). Specifically, EWER decreases from 4.16\% to 4.04\%
    for the \textbf{Small} list and from 5.61\% to 4.95\% for the \textbf{Large}
    list. These results demonstrate that entity filtering effectively enhances biasing
    performance, particularly when handling large and noisy biasing lists. Additionally,
    general transcription performance, as measured by WER, is also slightly
    improved—from 4.54\% to 4.46\% for the \textbf{Small} list and from 4.80\%
    to 4.71\% for the \textbf{Large} list—indicating that the entity filtering
    strategy not only improves entity recognition but also helps maintain
    overall transcription quality.

    \subsubsection{Comparison to Shallow Fusion}
    Shallow fusion is a widely adopted technique for enhancing contextual
    biasing performance in ASR systems. As shown in Table~\ref{tab:biasing}, we
    compare our proposed Prompt Biasing model with entity filtering (rows D1 and
    D2) against the baseline model with shallow fusion (rows B1 and B2) across
    different biasing list sizes. Our approach consistently outperforms shallow fusion,
    achieving a relative reduction in EWER of 30.7\% for the \textbf{Small}
    biasing list (D1 vs. B1) and 18.0\% for the \textbf{Large} biasing list (D2
    vs. B2). These results highlight the effectiveness of Prompt Biasing with entity
    filtering in leveraging contextual information to improve entity recognition
    accuracy. Additionally, our method maintains competitive overall transcription
    quality, with WERs of 4.46\% and 4.71\% for the \textbf{Small} and \textbf{Large}
    lists, respectively, compared to 4.67\% and 4.74\% for shallow fusion. This demonstrates
    that our approach not only enhances entity recognition but also preserves general
    ASR performance, even with large and noisy biasing lists.

    \subsubsection{Robustness Against Noise}
    Table~\ref{tab:robustness} presents the performance of the models on a 7.6M-word
    in-house dataset under both standard (no biasing list) and noisy biasing list
    conditions. The standard setting corresponds to conventional ASR without
    contextual information. Compared to the baseline, the Prompt Biasing model
    exhibits only a marginal increase in WER of 0.04\% in the absence of contextual
    information, demonstrating minimal impact on general ASR performance. Furthermore,
    when evaluated with a noisy biasing list, the Prompt Biasing model shows a negligible
    WER degradation of 0.06\%, indicating strong robustness to irrelevant or inaccurate
    contextual information.
    \begin{table}[t!]
        \centering
        \caption{Performance on in-house dataset for robustness benchmark.}
        \begin{tabular}{lcc}
            \hline
            Model          & Biasing List & WER(\%) \\
            \hline
            Baseline       & N/A          & 6.91    \\
            \hline
            Prompt Biasing & N/A          & 6.95    \\
            Prompt Biasing & Noisy        & 6.97    \\
            \hline
        \end{tabular}
        \label{tab:robustness}
    \end{table}
    \section{Conclusion and Future Work}
    In this paper, we presented a lightweight and effective prompt-based contextual
    biasing approach for Transformer-based ASR systems. By casting contextual
    biasing as a multi-task learning problem, our method enables seamless integration
    of contextual information without requiring any architectural modifications.
    We also introduced an efficient entity filtering mechanism that
    significantly enhances biasing performance, especially when dealing with large
    and noisy biasing lists. Experimental results show that our approach
    consistently outperforms shallow fusion techniques in both entity
    recognition and overall transcription accuracy. In future work, we plan to
    investigate the combination of Prompt Biasing with shallow fusion and
    explore the integration of large audio language models to further advance
    contextual ASR capabilities.
    \bibliographystyle{IEEEtran}
    \bibliography{mybib}
\end{document}